# Static and dynamic simulation in the classical two-dimensional anisotropic Heisenberg model

J. E. R. Costa and B. V. Costa

*Departamento de Física, Instituto de Ciências Exatas, Universidade Federal de Minas Gerais, Caixa Postal 702, 30161-970 Belo Horizonte, Brazil*

(Received 21 December 1995)

By using a simulated annealing approach, Monte Carlo and molecular-dynamics techniques we have studied static and dynamic behavior of the classical two-dimensional anisotropic Heisenberg model. We have obtained numerically that the vortex developed in such a model exhibit two different behaviors depending if the value of the anisotropy $\lambda$ lies below or above a critical value $\lambda_c$. The in-plane and out-of-plane correlation functions ($S^{xx}$ and $S^{zz}$) were obtained numerically for $\lambda<\lambda_c$ and $\lambda>\lambda_c$. We found that the out-of-plane dynamical correlation function exhibits a central peak for $\lambda>\lambda_c$ but not for $\lambda<\lambda_c$ at temperatures above $T_{KT}$. [S0163-1829(96)04626-7]

## I. INTRODUCTION

In the last ten years much attention has been dedicated to the study of the classical two-dimensional anisotropic Heisenberg model (CTDAHM). Such attention is grounded in the fact that a large variety of models may be mapped in the CTDAHM. Some examples are superfluid, superconducting films and roughening transitions.[1–9] The CTDAHM can be described by the Hamiltonian

$$H = -J \sum_{\langle i,j \rangle} (S_i^x S_j^x + S_i^y S_j^y + \lambda S_i^z S_j^z), \quad (1)$$

where $J>0$ is the exchange coupling constant which defines a ferromagnetic system and $\lambda$ is an anisotropy, $S^\alpha$ are classical spin components defined on the surface of a unit sphere, and $\langle i,j \rangle$ are to be understood as first neighbor indices in a square lattice. For $\lambda=0$ we obtain the so-called *XY* model (that should not be confused with the planar model, that has only two spin components) and for $\lambda=1$ the pure Heisenberg model. Thermodynamic properties of the Hamiltonian (1) are well understood in the limit $\lambda=0$ following the work of Berezinskii[10] and Kosterlitz and Thouless.[11] It has a phase transition of infinite order at temperature $T_{KT}$, named Kosterlitz-Thouless phase transition, with no long-range order, which is characterized by a vortex-anti-vortex unbinding.[12] A vortex (antivortex) is a topological excitation where spins on a closed path around the excitation precess by $2\pi$ ($-2\pi$) in the same direction. When $T_{KT}$ is reached from above the correlation length $\xi$ and magnetic susceptibility $\chi$ behave as

$$\xi \sim e^{b_\xi t^{-\nu}}, \quad \chi \sim e^{b_\chi t^{-\eta}},$$

where $t=(T-T_{KT})/T_{KT}$, with $b_\xi, b_\chi \sim 1$, $\nu=\frac{1}{2}$ and $\eta=\frac{1}{4}$. When the system goes through $T_{KT}$ a vortex-anti-vortex unbinding process occurs increasing the entropy in the system. For $\lambda<1$ we expect the same thermodynamic behavior since the system is in the same universality class as the *XY* model. The variation of $T_{KT}$ with $\lambda$ is experimentally important. Both analytical as well simulational results show that $T_{KT}$ depends weakly on $\lambda$, except for $\lambda \sim 1$ when $T_{KT} \to 0$.

The dynamical behavior of the *XY* model was studied theoretically with different predictions for the nature of the neutron-scattering function. Villain[13] analyzed the model in the low-*T* limit in the harmonic approximation. He found that the in-plane $S^{xx}$ correlation function behaves as

$$S^{xx}(q,\omega) \sim |\omega - \omega_q|^{-1+\eta/2}$$

with $\eta=1/4$ at $T_{KT}$, $\omega_q$ is the magnon frequency, and $S^{xx}(q,\omega)$ is obtained by Fourier transforming the space-time correlation function.

In a hydrodynamic description, without vortex contribution, Nelson and Fisher[14] found the in-plane correlation function

$$S^{xx}(q,\omega) \sim \frac{1}{q^{3-\eta}} \Psi\left(\frac{\omega}{q}\right),$$

where

$$\Psi(y) \sim \frac{1}{|1-y^2|^{1-\eta}}$$

around the spin-wave peak and

$$S^{xx}(q,\omega) \sim \omega^{\eta-3}$$

for large values of $\omega/q$.

Both, Villain and Nelson and Fisher predicted a narrow spin-wave peak to the out-of-plane correlation function $S^{zz}$. By performing a low-temperature calculation which includes out-of-plane contributions, Menezes *et al.*[15] found a spin-wave peak similar to that of Nelson and Fisher. In addition to the spin-wave peak they found a central logarithmically divergent peak.

At $T_{KT}$ vortex-anti-vortex pairs start to unbind, and vortices may diffuse through the system leading to a strong central peak, at the same time the stiffness jumps to zero meaning that the spin-wave peaks disappear.[14,16]

Mertens *et al.*[17] have proposed a phenomenological model to calculate the correlation function above $T_{KT}$. Their approach was based on a well succeeded ballistic approach to the one-dimensional soliton dynamics in magnetic spin chains.[18] They found a Lorentzian central peak for $S^{xx}$ and a Gaussian central peak for $S^{zz}$.

In a recent work, Evertz and Landau[19] using spin-dynamics techniques have performed a large-scale computer simulation of the dynamical behavior of the $XY$ model. They found an unexpected central peak in the $S^{xx}$ correlation function for temperatures well below $T_{KT}$, and their results are not adequately described by above theories.

From the experimental point of view, Wiesler *et al.*[21] studied a very anisotropic material that is expected to have a $XY$ behavior. For $T<T_{KT}$ they found spin-wave peaks but it is not clear if a central peak is present. Above $T_{KT}$ they found the expected central peak in the in-plane correlation function and the out-of-plane function exhibits damped spin waves. More recently Song[22] performed $^{89}$Y NMR experiments on a powder sample of type-II superconductor YBa$_2$Cu$_3$O$_{7-\delta}$ around the Kosterlitz-Thouless temperature in a magnetic field. In their experiment was observed only local vortex motion and diffusive behavior seems to be absent.

Since recent experimental works and numerical ones present results that are not in accordance with existent theories, mainly in aspects related to the central peak and vortex motion, more work to investigate this subject is justifiable. In fact, we found some results which are in agreement with Song's observation about vortex motion, that we present at the conclusion.

In this work we present some Monte Carlo and molecular-dynamics simulation in the CTDAHM defined by the Hamiltonian (1) for $\lambda \neq 0$. In Sec. II we discuss the effect of finite anisotropy to both vortex components, in-plane and out-of-plane. A limiting value $\lambda_c$ for the anisotropy is numerically obtained. For $\lambda < \lambda_c$ the most stable spin configuration is a planar one. For $\lambda > \lambda_c$ it develops a large out-of-plane $S^z$ component near the center of the vortex. The behavior of $S^z$ as a function of $\lambda$ is obtained. In Sec. III we calculate, by using Monte Carlo (MC) and molecular-dynamics simulation the in-plane correlation function $S^{xx}$ and out-of-plane $S^{zz}$, for two values of $\lambda$, $\lambda < \lambda_c$ and $\lambda > \lambda_c$. Finally in Sec. IV we present our conclusions pointing out the relevant aspects introduced by a finite anisotropy.

## II. STATIC VORTEX SOLUTIONS

In this section we discuss the static vortex solutions to the Hamiltonian (1) for arbitrary $0 \leq \lambda < 1$, firstly in the continuum limit and then we obtain numerical solutions to the discrete case.

The classical spin vector may be parametrized by the spherical angles $\Theta_n$ and $\Phi_n$ as

$$\vec{S}_n = (\cos\Theta_n \cos\Phi_n, \cos\Theta_n \sin\Phi_n, \sin\Theta_n). \quad (2)$$

In the continuum approximation for the Hamiltonian (1), $\Phi_n$ and $S_n^z = \sin\Theta_n$ constitute a pair of canonically conjugate variables, which allow us to write the equations of motion

$$\dot{\Theta}_n = \frac{\partial H/\partial \Phi_n}{\cos\Theta_n}, \quad \dot{\Phi}_n = \frac{\partial H/\partial \Theta_n}{\cos\Theta_n}. \quad (3)$$

If $S^\alpha$ has an expansion like

$$S^\alpha(x_n \pm a, y_n) = \sum_{k=0}^{\infty} \left( \pm a \frac{d}{dx_n} \right)^k S^\alpha(x_n, y_n), \quad (4)$$

where $a$ is the lattice constant, we may rewrite (3) in a continuum version as

$$\dot{\Theta} = 2J[\cos\Phi \nabla^2(\cos\Theta \sin\Phi) - \sin\Phi \nabla^2(\cos\Theta \cos\Phi)] \quad (5)$$

$$-\cos\Theta \dot{\Phi} = 2J\{\lambda \cos\Theta[\sin^2\Phi \nabla^2 \sin\Theta + \cos^2\Phi \nabla^2(\sin\Theta \cos\Phi)] - \sin\Theta[\sin\Phi \nabla^2(\cos\Theta \sin\Phi) + \cos\Phi \nabla^2(\cos\Theta \cos\Phi)]\}, \quad (6)$$

where we kept term up to order $a^2$ in the expansion.

Single static vortex solution may be obtained from Eqs. (5) and (6) with the appropriate boundary conditions[23]

$$S^\alpha(x,y) = S^\alpha(-x,y), \quad \lim y \to \pm \infty,$$

$$S^\alpha(x,y) = S^\alpha(x,-y), \quad \lim x \to \pm \infty, \quad (7)$$

where $\alpha = x, y$. One solution can be readily seen as

$$\Theta_0 = 0 \quad \text{and} \quad \Phi_0 = \arctan\frac{y}{x}, \quad (8)$$

which describes an in-plane vortex (KT).

We expect that the expression given by (8) should be a stable solution for moderate values of the anisotropy $\lambda$ since as long as $\lambda$ grows a smaller energy configuration may be achieved if the $S^z$ component develops a nonzero value near the vortex center. Unfortunately a complete analytical solution with $\Theta \neq 0$ is not available so far. However, if we consider the limits $r \to 0$ and $r \to \infty$, where $r$ is the distance from the center of the vortex, we may write approximate solutions:

$$\Phi = \Phi_0, \quad (9)$$

$$\sin\Theta = \begin{cases} 1 - Ar^2, & \text{if } r \to 0 \\ B\exp\left\{-2\left(\frac{1-\lambda}{\lambda}\right)^{1/2} r\right\}, & \text{if } r \to \infty \end{cases}$$

We do not really expect that the expressions given by (8) and (9) are good solutions for the discrete case in the limit $r \to 0$ since variations on $\Theta$ should be stronger there. It is straightforward to calculate the contributions to the energy in the continuum limit due to both configurations, Eqs. (8) and (9), they are dominated by a $\ln r$ term. We will see below that this behavior persists up to values of $\lambda$ quite near $\lambda = 1$. At the Heisenberg limit the energy has a completely different

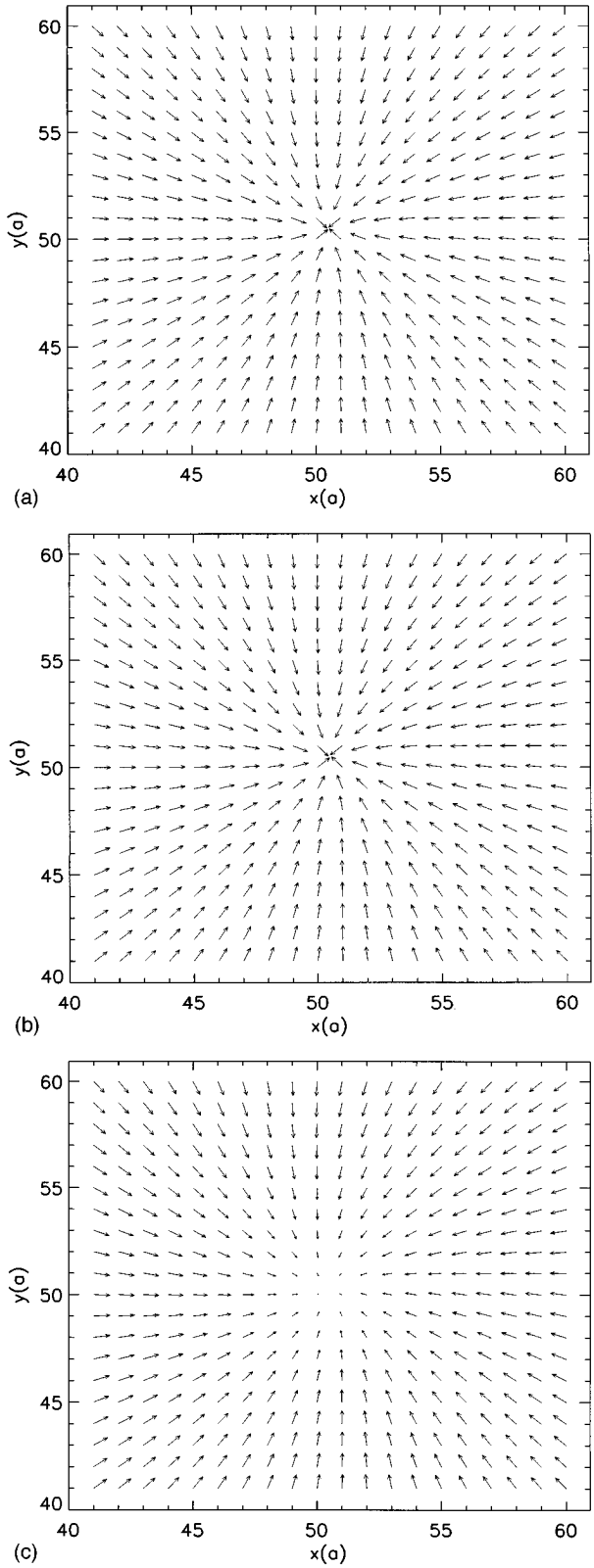

FIG. 1. In-plane spin components in a square lattice of linear length $L=100$. Only the central region with 400 spins are shown for a better visualization. One can see the presence of only one vortice due to boundary conditions and $T_{min}=10^{-5}$. The values of anisotropy are (a) $\lambda=0.700$; (b) $\lambda=0.710$; (c) $\lambda=0.990$.

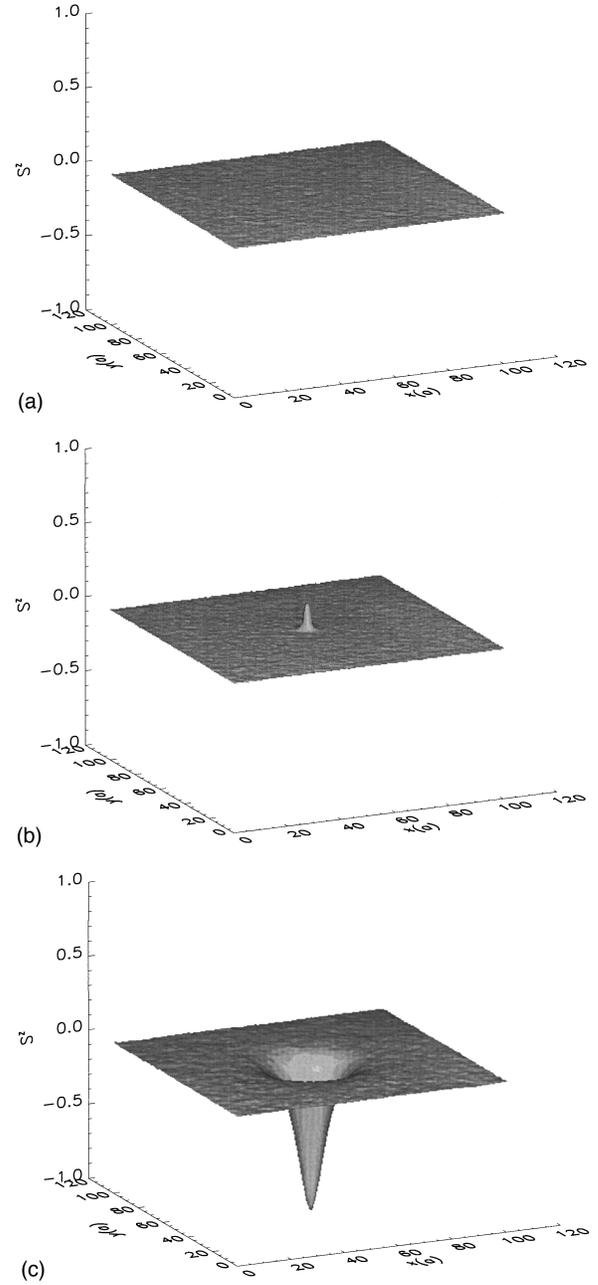

FIG. 2. Out-of-plane spin components in a square lattice of linear length $L=100$. Parameters are the same as in Fig. 1.

behavior, the relevant excitations become instantons rather than vortices which energy is a constant.[20]

In order to solve the discrete equations of motion given by (3) in the static case, we use a simulated annealing approach, which was shown to be quite powerful in determining absolute minimum in spin-glass models.[24] We minimize the Hamiltonian (1) using diagonally antiperiodic boundary conditions to the $x$ and $y$ spin components and diagonally periodic one to the $z$ component

$$S_{i,0}^{\alpha}=kS_{L-i,L}^{\alpha}, \qquad (10)$$

$$S_{0,j}^{\alpha}=kS_{L,L-j}^{\alpha},$$

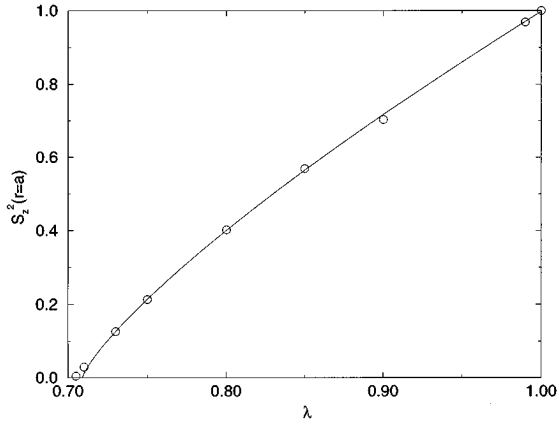

FIG. 3. Out-of-plane squared component ($S_z^2$) as a function of $\lambda$. Circles are simulation points (from numerical vortex solution at $T \approx 0$) and the solid line is the fit using $(S^z)^2 \sim (\lambda - \lambda_c)^\nu$.

where $k = -1$ if $\alpha = x,y$ and $k = 1$ if $\alpha = z$ and $0 \leq i,j \leq L$. These boundary conditions are enough to create an odd number of vortex (antivortex) in the system and the ground state has only one vortex (antivortex), so that we can find numerically the stable vortex solution (in-plane or out-of-plane) for each value of $\lambda$ anisotropy.

We started the iteration by using the exact continuum vortex solution given by (8) in a square lattice of linear size $L = 100$. In some cases we have used $L = 400$ with no significant change in the final results. The iterative simulated annealing process is implemented starting at temperature $T_{init} = 0.1$ until the minimum $T_{min} = 10^{-5}$. Steps in temperature, $\Delta T$, are chosen so that the acceptance rate is maintained in 50%.

Results for $S^{x,y}$ and $S^z$ are shown in Figs. 1 and 2 for some values of the anisotropy $\lambda$. Two types of behavior are quite clear. There is a region of $\lambda$ where the stable solution is

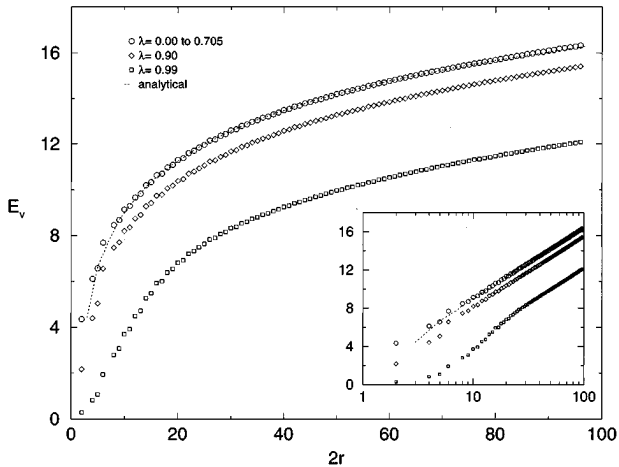

FIG. 4. Vortex energy (measured in units of $J$) as a function of the vortex diameter $2r$. (r is measured in units of lattice spacing.) Energy curves for $\lambda = 0.710$, $\lambda = 0.900$ $\lambda = 0.990$ are shown as circles, diamonds, and squares, respectively. The dotted line comes from Eq. (8). The inset shows a log-linear plot of energy as a function of ln$2r$.

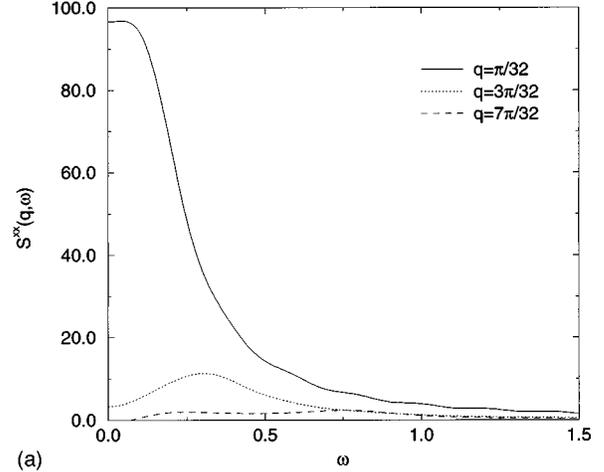

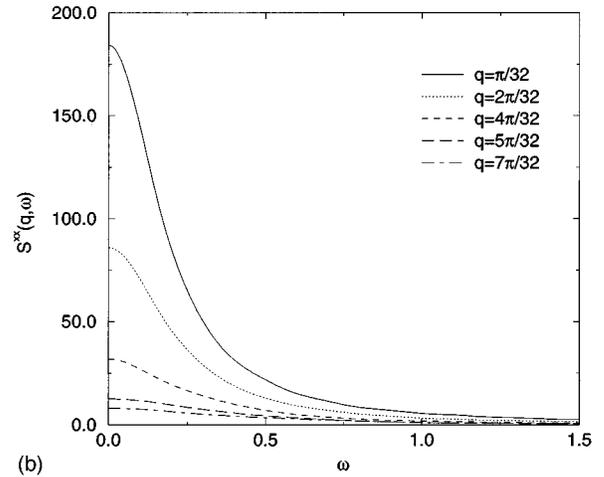

FIG. 5. In-plane correlation function $S^{xx}(q,\omega)$ as a function of $\omega$ for $\lambda = 0.5$ and (a) $T = 0.60J/k_0$ ; (b) $T = 0.80J/k_0$. Values for $q$ are shown in the inset.

$S^z = 0$ and another region where $S^z \neq 0$ near the vortex center. We observe that for $\lambda \approx \lambda_c$ the $S^z$ component is appreciable only inside a small region near the vortex core. As long as $\lambda$ increases, $S^z$ becomes larger and the vortex core grows. Because of the $S^z$ symmetry $+S^z$ and $-S^z$ are equivalent solutions. In order to determine the critical value of $\lambda$ we have obtained a series of solutions for $S^z$ and $S^{x,y}$ for different values of $\lambda$. By measuring the $S^z$ component at $x = y = a$ (where $a$ is a lattice constant) we determined where it goes to zero. A plot of such results is shown in Fig. 3. The $S^z$ component goes to zero at $\lambda_c \simeq 0.709 \pm 0.001$. The behavior of $S^z$ as a function of $\lambda$ is well described by a function $(S^z)^2 \sim (\lambda - \lambda_c)^\nu$ with $\nu = 0.785 \pm 0.004$. Of course the existence of a $\lambda_c$ does not mean that the system undergoes a phase transition, but just that the vortex develops an out-of-plane component from this value of $\lambda$. We have also calculated the energy as a function of the distance to the center of the vortex for some values of $\lambda$. Energy curves obtained by simulation are shown in Fig. 4 as circles, diamonds, and squares for $\lambda = 0.710$, $0.900$, and $0.990$, respectively. The dotted line comes from the exact continuous solution given by Eq. (8). The inset shows a log-linear plot of energy as a function of ln$2r$. The deviation from the logarithmic behavior is clear.

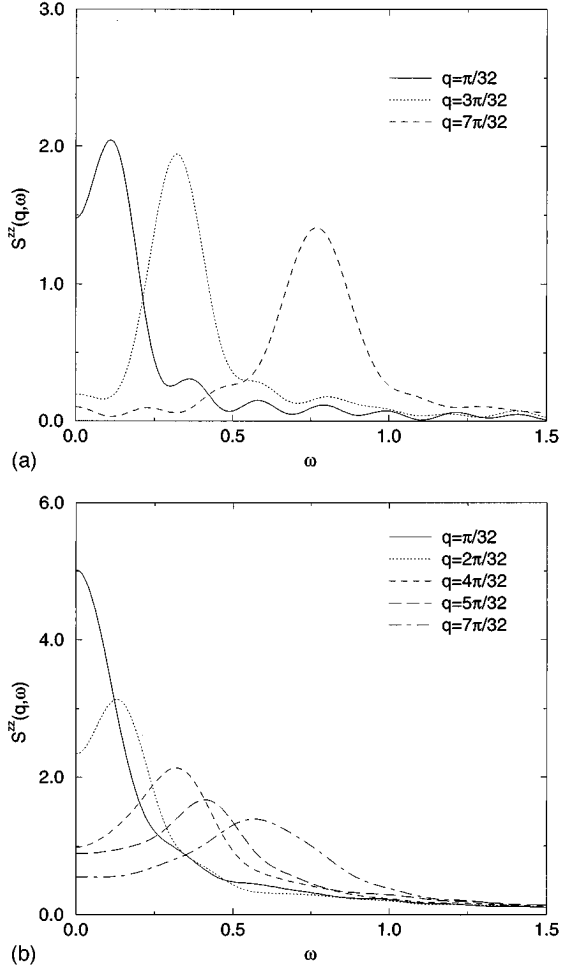

FIG. 6. Out-of-plane correlation function $S^{zz}(q,\omega)$ as a function of $\omega$ for $\lambda=0.5$ and (a) $T=0.60J/k_0$ ; (b)$T=0.80J/k_0$. Values for $q$ are shown in the inset.

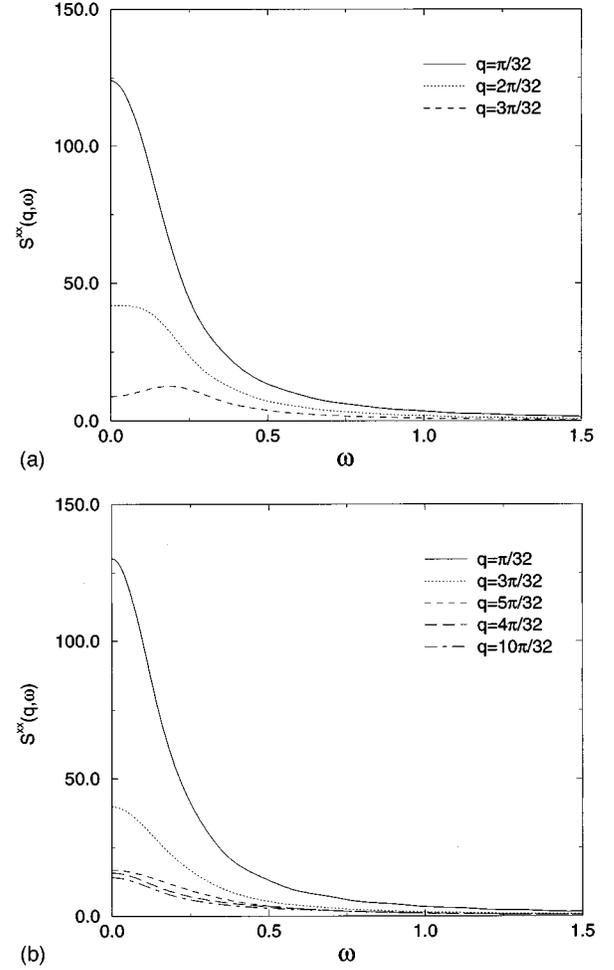

FIG. 7. In-plane correlation function $S^{xx}(q,\omega)$ as a function of $\omega$ for $\lambda=0.8$ and (a) $T=0.60J/k_0$; (b) $T=0.80J/k_0$. Values for $q$ are shown in the inset.

Now a natural question arises, how a finite anisotropy changes the dynamical correlation functions $S^{xx}$ and $S^{zz}$? Because the in-plane symmetry is not changed we do not expect any drastic change in $S^{xx}$. However, since for $\lambda > \lambda_c$ the most stable vortex solution is for $S^z \neq 0$ the development of a central peak for $S^{zz}$ will not be surprising. In the next section we numerically calculate both $S^{xx}$ and $S^{zz}$.

### III. DYNAMICAL CORRELATION FUNCTIONS

In the last section we observed a drastic change in the $S^z$ spin component when the anisotropy $\lambda$ exceeds $\lambda_c$. Because the in-plane symmetry is not changed when $\lambda$ goes through $\lambda_c$ we do not expect any drastic change in $S^{xx}$. However, since the most stable configuration changes suddenly from $S^z=0$ to a nonzero value the development of a central peak in the $S^{zz}$ correlation function will not be surprising.

In this section we present Monte Carlo–molecular-dynamic simulation results we carried out to obtain the correlation function $S^{xx}$ and $S^{zz}$ for two values of $\lambda$, below ($\lambda=0.50$) and above ($\lambda=0.80$) the critical anisotropy. Our simulations were done on a $64\times 64$ square lattice with periodic boundary conditions at temperature $T=0.60$ and $T=0.80$ in units of $J/k_0$ which are below and above the Kosterlitz-Thouless temperature $T_{KT}$.

Equilibrium configurations were created at each temperature using a Monte Carlo method which combines cluster updates of in-plane spin components[25] with Metropolis reorientation. After each single cluster update, two Metropolis sweeps were performed. The cluster update is essential at the low-temperature region, since the critical slowing down is severe and it should not be possible to achieve thermodynamic equilibrium in a reasonable computer time using only the Metropolis algorithm. We have used in our simulation 200 independent configurations discarding the first 5000 hybrid sweeps for equilibration.

Starting with each equilibrated configuration, the time spin evolution was determined from the coupled equations of motion for each spin[26]

$$\frac{d}{dt}\vec{S}_{i,j} = \vec{S}_{i,j} \times \vec{V}_{i,j}, \qquad (11)$$

where

$$\vec{V} = J\sum_{\alpha} (S^{\alpha}_{i-1,j} + S^{\alpha}_{i,j-1} + S^{\alpha}_{i+1,j} + S^{\alpha}_{i,j+1})\hat{e}_{\alpha}$$

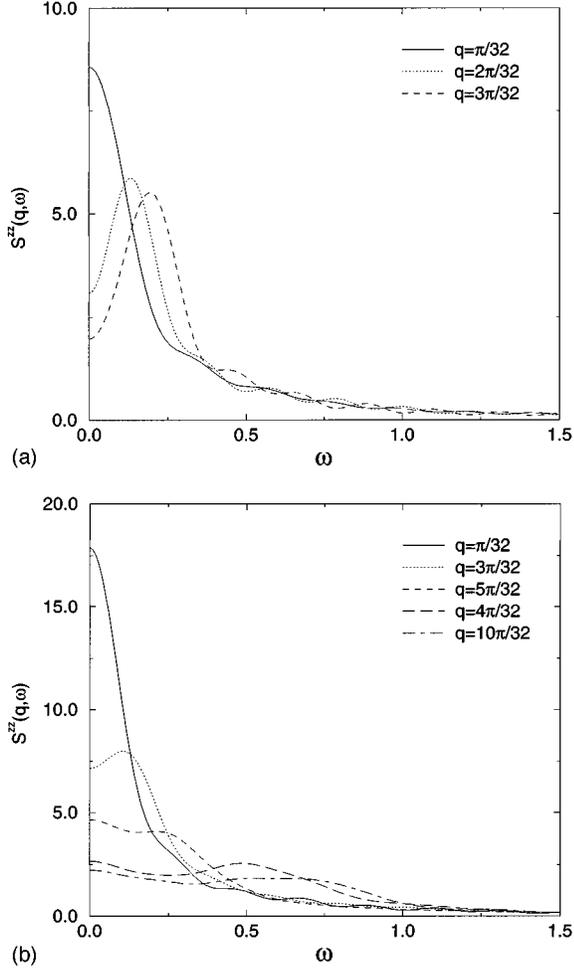

FIG. 8. Out-of-plane correlation function $S^{zz}(q,\omega)$ as a function of $\omega$ for $\lambda=0.8$ and (a) $T=0.60J/k_0$; (b) $T=0.80J/k_0$. Values for $q$ are shown in the inset.

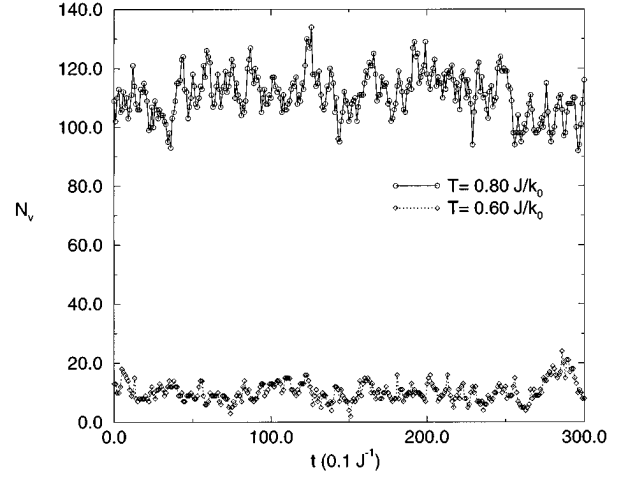

FIG. 9. Number of vortices ($N_v$) as a function of time $[t(0.1J^{-1})]$.

and $\alpha=x,y,z$, $\hat{e}_x$ and $\hat{e}_y$ are unit vectors in the $x$ and $y$ direction, respectively. Equation (11) was numerically integrated by using a fourth-order predictor-corrector method[27] with a time step of $\delta t=0.025J^{-1}$. The maximum integration time was $t_{max}=30J^{-1}$. A few runs with $t_{max}=60J^{-1}$ were done with the same results for our purpose, giving the same physical results. The numerical integration stability is checked out verifying that the constants of motion (energy and $z$ magnetization) remain constants with a relative variation of less than $10^{-6}$ after 1200 time steps. To obtain $S^{\alpha\alpha}(q,\omega)$ we first calculated the space-time correlation functions, $S^{\alpha\alpha}(i-j,t)$ as

$$\langle S_i^\alpha(0,0) S_j^\alpha(r,t)\rangle = \frac{1}{N} \sum_{i=1}^{N} \sum_{j=1}^{N} S_i^\alpha(0) S_j^\alpha(t) \quad (12)$$

for time steps of size $\Delta t=0.1J^{-1}$ up to $0.9t_{max}$ and finally averaging over all configuration.

By Fourier transformation in space and time we have obtained the neutron-scattering function $S^{\alpha\alpha}(q,\omega)$. We restrict ourselves to momenta $\vec{q}=(q,0)$ and $(0,q)$ with $q$ given by

$$q=n\frac{2\pi}{L}, \quad n=1,2,\ldots,L.$$

Since these two directions are equivalent we averaged them together to get better statistical accuracy. The frequency resolution of our results is determined by the time integration cutoff ($=0.9t_{max}$) which introduces oscillations into $S^{\alpha\alpha}(q,\omega)$. To reduce the cutoff effects we introduced Gaussian spatial and temporal functions[28] replacing $S^{\alpha\alpha}(r,t)$ by

$$S^{\alpha\alpha}(r,t) e^{-(1/2)(t\delta\omega)^2} e^{-(1/2)(r\delta q)^2}$$

to compute $S^{\alpha\alpha}(q,\omega)$. Cutoff parameters are $\Delta\omega=0.05$ and $\Delta q=0.05$.

Figures 5(a) and 5(b) show $S^{xx}(q,\omega)$ for $T=0.6$ and 0.8, and $\lambda=0.5$. There is no substantial difference from those results obtained by Evertz and Landau in Ref. 19 to the case $\lambda=0$. At low $T$ we have only spin-wave peaks and $T>T_{KT}$ only central peaks are displaced. The out-of-plane $S^{zz}$ correlation function is shown in Fig. 6 to the same parameters and only spin-wave peaks are observed. The interesting behavior comes up when we go through $\lambda_c$. In Figs. 7 and 8 we show the in-plane and out-of-plane correlation functions, respectively. To $S^{xx}$ we observed the same qualitative behavior for $\lambda<\lambda_c$, however for $S^{zz}$ a very clear central peak is developed for $T>T_{KT}$. As commented before the source of such a central peak seems to lie on the vortex structure developed for $\lambda>\lambda_c$.

## IV. CONCLUSIONS

We have obtained numerically that the vortex developed in the CTDAHM exhibit very different behavior depending if the value of the anisotropy $\lambda$ lies below or above the critical value $\lambda_c$. For $\lambda<\lambda_c$ the spin components lie preferentially in the $XY$ plane, while for $\lambda>\lambda_c$ the most stable configuration develops an out-of-plane component that grows with $\lambda$. We have shown that the out-of-plane dynamical correlation function has a central peak for $\lambda>\lambda_c$ but not for $\lambda<\lambda_c$. Theories developed so far did not describe correctly the correlation function as discussed in Refs. 19, 21, and 22. In an earlier work (Costa *et. al*[29]) suggested that central peak might be due to a vortex-anti-vortex creation annihilation process. We have calculated in our simulations the fluctuation of the number of vortex with time for all configurations, anisotropies and temperatures. Figure 9 is a typical plot of the number of vortex as a function of time. Below and above $T_{KT}$ the fluctuation of the number is very strong. Pairs may annihilate at the position $\vec{r}$ on time $t$, reappearing at $\vec{r}'$ on $t'$. This process may introduce the dynamics to give the central peaks. This is in accordance with the NMR results of Song in Ref. 22 who found only local vortex motion in his measurements and with the central peak for $T<T_{KT}$ found by Evertz and Landau[19] in the in-plane correlation function. An analytical calculation using a Master equation approach in order to incorporate the creation-annihilation process is now in progress.

## ACKNOWLEDGMENTS

We thank FAPEMIG and CNPq for financial support. Part of our computer simulations were carried out on the Cray YMP at CESUP (UFRGS).